# Novel crossbar array of silicon nitride resistive memories on SOI enables memristor rationed logic

N. Vasileiadis [a,b], A. Mavropoulis [a], I. Karafyllidis [b], G. Ch. Sirakoulis [b], P. Dimitrakis [a]

[a] *Institute of Nanoscience and Nanotechnology, NCSR "Demokritos", Ag. Paraskevi 15341, Greece*
[b] *Department of Electrical and Computer Engineering, Democritus University of Thrace, Xanthi 67100, Greece*

## ABSTRACT

In this work, the fabrication of crossbar arrays of silicon nitride resistive memories on silicon-on-insulator substrate and their utilization to realize multi-rationed logic circuits are presented. Typical electrical charac-terization of the memristors revealed their ability of multi-state operation by the presence of 12 well separated resistance levels. Through a dedicated modeling and fitting procedure of these levels, a reconfigurable logic based on memristor rationed logic scheme is designed and a crossbar integration methodology was proposed. Finally, several circuitry aspects were simulated in SPICE with a silicon nitride SOI crossbar array calibrated model and power optimization prospects were discussed.

## 1. Introduction

Resistive memories (RRAMs) are one of the most promising nonvolatile memory alternatives [1]. RRAM cells are ideal for use in crossbar architecture, which can provide the smallest memory cell conceivable [2]. For in-memory and neuromorphic computing accelerators [3] and particularly for low power mobile IoT edge computing hardware [4,5], such an approach is highly intriguing.

Memristor rationed logic (MRL) on the other hand is an upcoming technology that promises to boost the RRAM evolution through a combination of CMOS and memristor technology [6]. Integrated circuit designers typically use MRL to replace as many transistors as possible with memristors, resulting in relatively dense and scalable systems. Many initiatives to rethink the fundamental digital building blocks have been made in the literature based on these memristive logic design patterns [7]. However, state-of-the-art research in this domain is not directly optimized to fit with different RRAM technologies. Moreover, the majority of these design approaches tends to operate RRAM devices at their maximum HRS/LRS ratio. This operation scheme reduces their endurance expectancy [8–12] due to the frequent resistance switch at each computational step.

In our last research work, we have demonstrated the superiority of SOI on silicon nitride RRAM MIS devices [13]. Here, the implementation of a 6 × 6 crossbar (Xbar) array of one-resistor (1R) $SiN_x$ RRAM cells, with 300 nm lower dimension, upon an SOI substrate is presented. Through DC characterization and a custom-made fitting model the existence of 12 separable resistance states was established, which then were used for the calibration of the AND/OR gates of an MRL circuitry. Finally, the consumed power and the potentiality for low power applications were estimated and discussed respectively.

The rest of the paper is organized as follows. In Section 2, the fabrication method, the DC characterization and the multi-level-cell operation are presented accordingly. In Section 3, the reconfigurable circuitry is analyzed while several MRL circuit architecture considerations are discussed. Finally, Section 4 concludes the presented work.

## 2. Nanostructure fabrication and characterization

### 2.1. Fabrication method

A typical 6 × 6 1R crossbar structure fabricated on SOI substrate (100 nm p-Si and 200 nm buried oxide) is shown in Fig. 1. Firstly, the SOI was implanted with Phosphorus ions ($1 \times 10^{15}$ cm$^{-2}$ at 40 keV) through a 20 nm $SiO_2$ (hard mask) layer with a subsequent rapid thermal annealing at 1050 °C for 20 s. Next, a 7 nm $SiN_x$ layer was deposited by LPCVD at 810 °C, using ammonia ($NH_3$) and dichlorosilane ($SiCl_2H_2$) gas precursors [14]. An etching step was performed to create 1 μm wide fins of n$^{++}$-Si/SiN$_x$ stack where Si is acting as bottom electrode (crossbar columns) and LPCVD TEOS oxide was deposited to mutually isolate them (field oxide). Finally, 300 × 300 nm windows were opened on TEOS above the fins and top-electrode (TE) metal stripes and bottom-electrodes (BE) contact pads were defined by electron beam lithography for metal lift-off. TE metallization comprises a sputtered 30 nm Cu layer covered by 30 nm Pt to prevent ambient oxidation of Cu, while BE contact pads were formed by Al. The DC I-V characteristics of the fabricated devices were measured at room temperature using the HP4155A and Tektronix 4200A with pulse and amplification units.

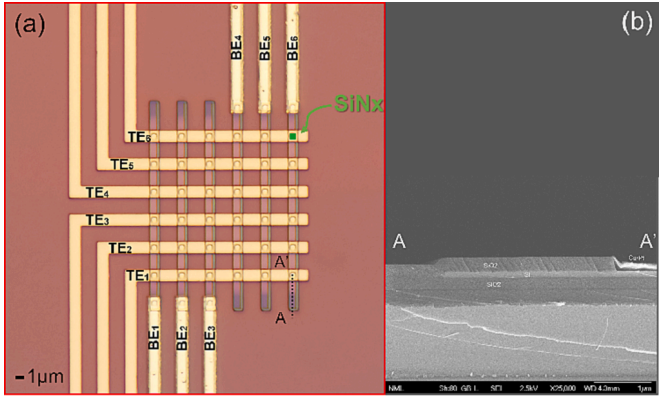

Fig. 1. (a) Optical microscope image of the fabricated 6 × 6 1R crossbar array structure on SOI. (b) SEM cross section micro-image across the AA' dashed line, as defined in (a).

### 2.2. Optical microscopy and SEM characterization

Optical microscope image of the crossbar circuit is shown in Fig. 1. The entire arrangement of the crossbar architecture together with the cells area are observed clearly. Also, in Fig. 1(b), a cross section SEM micro-image of the structure across the AA' dotted line (Fig. 1(a)) is shown. The buried oxide, the TEOS isolation layer and the footprint of Si layer BE with Cu metal layer (TE) are clearly observed through the different grey-scale contrast. The thin SiN$_x$ layer between TE and BE is also visible after careful observation.

### 2.3. DC characterization

In Fig. 2, a set of DC I-V sweeps for two SiN$_x$ resistive memory cells with different dimensions are shown. The nominal programming conditions encompassed double voltage sweeps within the range of $V_{TE} \in [-6V, 6V]$ and $I_{CC}$ [SET, RESET] = [10μA, 5 mA], respectively, performed over 30 repetitive cycles on both fresh devices (300 nm and 500 nm). The resistance switching mechanism in the tested memory cells is attributed to the presence of native traps created by the nitrogen deficiency and their subsequent density enhancement under the applied electric field. Evidently, during the I-V sweeps, significant current fluctuations are observed due to the unhindered electron exchange between the Si (BE) and the SiN$_x$ boarder traps [5,13,14]. It has been shown [5,14] that the presence of a thin tunneling oxide layer (2 nm) is suitable to mitigate this process and diminish the current fluctuations in I-V sweeps. The switching thresholds variability for the two examined sizes are shown in Fig. 3(a). The SET and RESET threshold voltage distributions of 1R cells in the crossbar exhibited greater variability compared to previous works with bulk and SOI single cell devices [13,14]. This is mainly attributed to the TE and BE resistance variations in the crossbar. However, the variability of both switching thresholds seem to increase when the size of the devices decreases; nevertheless, and beyond that, in both cases, we can distinguish discrete resistive levels, as depicted in the analysis provided below.

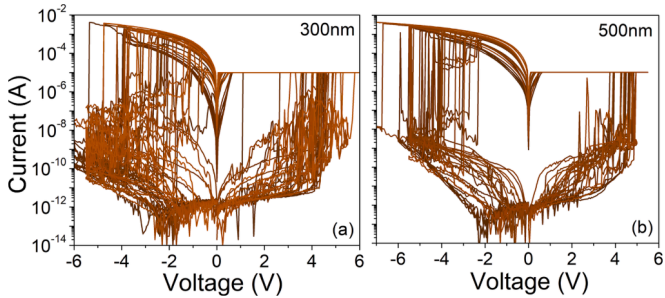

Fig. 2. 30 consecutive SET/RESET cycles for two resistive memory cells on SOI with dimension (a) 300 nm and (b) 500 nm, respectively.

### 2.4. Multi-Level-Cell (MLC) operation

Fig. 3(b) shows 12 well distinguishable high conductance states, which were extracted from the I-V curves shown in Fig. 2. In the initial part of the negative voltage sweeps, the SET process is completed, and the resistance filament is stabilized. The variabilities in conductance states at −0.8 V for each device over 30 voltage sweep cycles are visually presented in Fig. 3(c). It is noteworthy that the 300 nm devices exhibited a 55.8% broader range of programmed conductance states compared to the 500 nm devices, all measured at −0.8 V. After discarding some I-V's exhibiting RESET switching above −1V (outliers), the rest are displayed with opacity in Fig. 3(b). The 12 well separate low-resistance states which were extracted are marked with a different color (blue). Highly accurate fitting in the interval [−1.1 V, −0.4 V] was achieved using Eq. (1). This relation is a modified version of the one proposed by Strachan et al. [15] for modeling of tantalum oxide memristors. Unfortunately, Eq.(1) was not possible to fit the I-V curves up to 0 V due to some peculiarity in their curvature, and therefore our analysis is restricted in [−1.1 V, −0.4 V]. The inset of Fig. 3(b) shows the equivalent circuit which was used in place of the SiN$_x$ memristor in SPICE simulations. The circuit comprises a simple behavioral current source that generates a current with respect to Eq. (1), in parallel with a capacitance representing the memristor self-capacitance, which has been measured to be less than 10pF for the 300 nm size devices.

$$I = \alpha^* ||V|^{n_1}|^* e^{||V|^{n_2}|}, n_1, n_2 \in R \qquad (1)$$

## 3. Reconfigurable logic prospects

### 3.1. Reconfigurable memristive circuit structure

Fig. 4(a) shows a column of parallel connected SiN$_x$ memristors at a common node $V_L$ with a resistor $R_L$. This circuit is a reconfigurable structure as it is possible to vary the conductance state of each memristor resulting in change of the total current passes the node $V_L$ regulating the alternation of the voltage in the common node $V_L$. Thus, biasing each individual branch connected to the $V_L$ node and depending on the resistance level that the corresponding memristor is set, we can control the voltage at this node. For each new input that we want to connect to this reconfigurable structure, we should use an additional memristor. Consequently, we adopt the logic that is shown in the truth table of Fig. 4 (a), where logic 1 (True) and logic 0 (False) are represented by 1 V and Non-Contact (NC) respectively. This is because the connection of an input-potential lower than the other inputs would not produce the desired effect of the gradual increase of the potential at the $V_L$ node, which is a critical factor in the operation of this circuit. It is then sufficient to connect this circuit to an appropriately designed CMOS inverter, which will operate at a predetermined $V_{th}$ threshold. This reconfigurable SiN$_x$ memristive structure along with the CMOS inverter will form the SiN$_x$ MRL AND/OR gates.

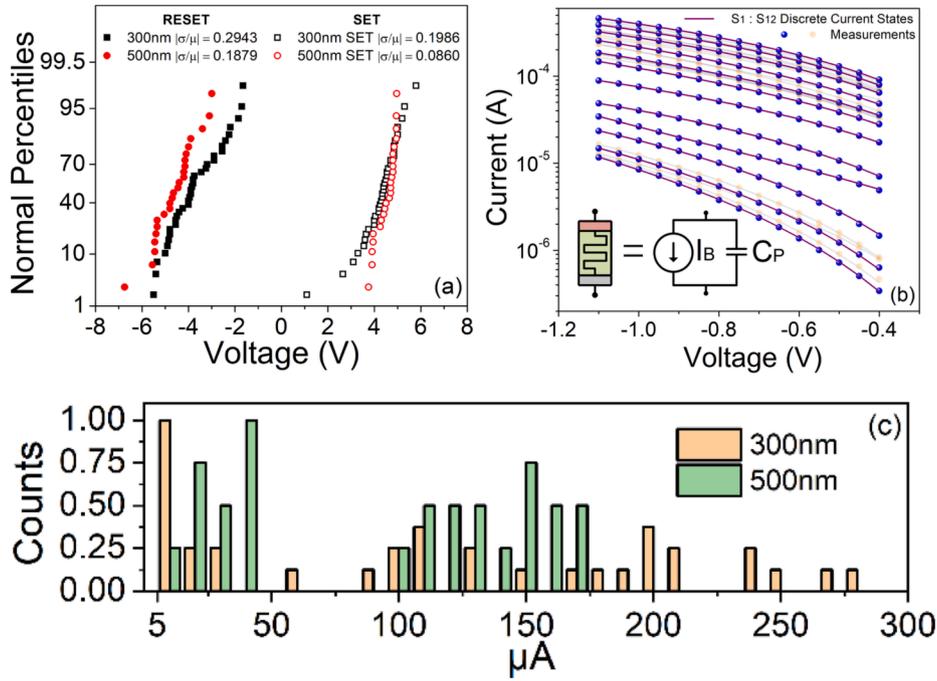

**Fig. 3.** (a) Switching thresholds variabilities, (b) fitting over extracted *I-V* characteristics from Fig. 2. Inset represents the equivalent SiNx RRAM SPICE model and (c) normalized counts of conductance states at −0.8 V for all 30 cycles of both devices.

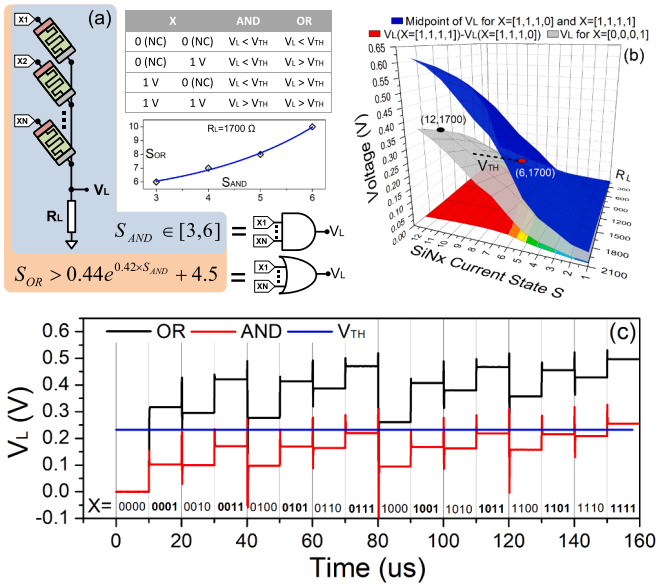

**Fig. 4.** (a) The proposed unconventional computing scheme based on a reconfigurable circuit of parallel connected SiN$_x$ devices on a common resistance node $V_L$ along with the conditions that the conductance states S must be satisfied by each gate instance. (b) Plot used in the analysis of selecting optimal parameters of the reconfigurable circuit based on the required criteria (4-input example). (c) SPICE simulation results of the variation of the voltage $V_L$ for 4-input MRL AND/OR gates.

Consequently, a methodology for the calculation of this critical threshold $V_{TH}$ should be determined. Ideally, in order to reduce the complexity of the proposed MRL, the same threshold $V_{TH}$ for both the AND and OR gates should be used so that the CMOS inverter can be designed once. Also, the resistor $R_L$ is desired to have a constant value. In this way, the design is simplified allowing to proceed to larger scale integrations. So, the question we have to answer is: *For the given range of inputs and conductance of the memristors, what resistance $R_L$ and what conductance states S should we choose for the memristors, so that the AND/OR gates can be implemented simultaneously with a unique threshold $V_{TH}$ and unique node resistance $R_L$, while at the same time ensuring the maximum separability between the voltages $V_L$ around the threshold $V_{TH}$?* An approach to this problem is demonstrated in the example below, where we created Fig. 4(b) using 4 SiN$_x$ memristors and the 12 discrete conductance states (see Fig. 3(b)) with a range of resistor $R_L = [300, 2100]$. The blue surface represents the midpoint variation of voltages $V_L$ for inputs $X = [1,1,1, NC]$ and $X = [1,1,1,1]$ and for various resistances $R_L$ and conductance states $S$. The answer is given by Fig. 4(b) where SPICE simulations revealed that the previous criteria are satisfied when $R_L = 1700\ \Omega$ and $[S = 6, S > 10]$ for the [AND, OR] gates, respectively. The colorful curve represents the difference of $V_L$ for $X = [1,1,1,NC]$ and $X = [1,1,1,1]$, while the gray curve represents the $V_L$ voltage for $X = [NC,NC,NC,1]$. Fig. 4(c) shows the result of the SPICE simulation for 4 SiN$_x$ memristors that in the first case implement an 4-input OR gate with $[R_L, S] = [1700, 12]$ while in the second an AND gate with $[R_L, S] = [1700, 6]$. The common threshold $V_{TH}$ that can effectively separate $V_L$ voltages is at 232 mV while the distance of $V_L$ values for $X = [1,1,1,NC]$ and $X = [1,1,1,1]$ is at 47 mV. This solution will be generalized for more inputs in the next paragraph 3.2.

Looking more carefully at Fig. 4(b), we could contact an analysis regarding the robustness of the proposed unconventional computing scheme. Keeping the criteria we defined in the previous paragraph (same threshold $V_{th}$ and unique node resistance $R_L$), for a fixed resistance $R_L = 1700\ \Omega$ and regarding the AND gate, the conductance states of the SiN$_x$ devices should belong to the range $S_{AND} = [3,6] \approx [18, 75]$ μA and this because, for conductance states $S < 3$, the separability of the outputs for $X = [1,1,1,NC]$ and $X = [1,1,1,1]$ falls below 20 mV. While for conductance states $S > 6$, there is no state $S > 12$ that allows for the implementation of the OR gate while respecting the initial condition for a same threshold $V_{th}$. Also, for each $S_{AND} \in [3,6]$, it should be $S_{OR} > 0.44e^{0.42 \times S_{AND}} + 4.5$, as shown by the exponential fitting to the data (inset plot Fig. 4(a): $f(x) = 0.44e^{0.42x} + 4.5$) extracted from Fig. 4 (b). In the most extreme case where $S_{AND} = 6$ it should be $S_{OR} \in [10,12] \approx [275, 397]$ μA. Therefore, we observe that in both cases of reconfigurable gates, we have a large margin of conductance state shift tolerance greater than 50 μA. This shifting of course does not come without any cost. In the case of the AND gate, the reduction of separability of the outputs down to 20 mV for a lower conductance state, requires more precise design of the CMOS output inverter in terms of non-linearity and noise interference.

### 3.2. MRL architecture considerations

The previous analysis should be extended to the case of a crossbar, where the intermediate resistances as well as the increasing number of inputs can affect the previous considerations. Fig. 5 shows a drawing of the MRL gate as it would be integrated into a crossbar arrangement. The Fig. 5 also presents the intermediate resistances in the horizontal ($R_h$) and vertical ($R_V$) connections of the crossbar. In our case, a resistivity of $\rho = 16.9\mu\Omega\cdot cm$ was measured for the horizontal Cu strips and $\rho = 1$ m$\Omega\cdot$cm for the vertical silicon strips, which then gave $R_H = 6.3\ \Omega$ and $R_V = 300\ \Omega$. The same figure also shows a pair of inverters that should be connected under each column of the crossbar. Their number increases according to the number of columns. The first inverter should meet the characteristics analyzed in the previous paragraph, while the second should also function as a tri-state buffer in order to ensure smoother interconnection with subsequent gates. The inverter pair also offers the advantage of a second boost stage in case the first inverter cannot deliver the required output. The design of this pair of CMOS inverters, however, is not the subject of this paper and will be considered in a subsequent work.

Fig. 6 shows some results of SPICE simulations by varying the number of inputs of the MRL gates. Simulations were performed both taking into account the intermediate resistances of the crossbar and without intermediate resistances. Also, for the sake of comparison, a crossbar with simple resistors instead of memristor was simulated. From the results is evident that as the number of inputs increases, the exponential behavior of the memristor's conductivity seems to "compress" more the voltage $V_L$ compared to the case of the simple resistance, while in the case where there are also intermediate resistances, saturation is observed when the number of 10 inputs is approached. This is reasonable because for the same number of inputs these voltage drops over the intermediate resistors leads to a smaller voltage drop over the memristor itself and this in turn to an even smaller current due to the exponential behavior of the memristor's conductivity. A red dotted line has also been placed in the Fig. 6, which represents the maximum common threshold $V_{th}$ based on which the MRL gates can be implemented as we discussed in the previous paragraph 3.1. At first sight the intermediate resistors seem to help provided that 10 or more inputs are required in order for the separation to stop being possible with a common threshold. However, the separability in this case seems to decrease faster and in cases where the design capabilities of the CMOS inverter or the noise levels of the system do not allow it, we are forced to select fewer inputs where the separability is greater. Especially for the case of SiN$_x$ SOI crossbar, 6-input gates seem to be the ideal case while keeping the distance of $V_L$ for $X_5$ and $X_6$ at 20 mV. The size of the number of columns does not seem to have an effect in our case, as we have a line resistance of 2 orders of magnitude lower than the rest of the resistances in the circuit. Larger MRL gates can be achieved by combining more smaller crossbars using the segmented architecture of crossbars [16,17], i.e., several smaller crossbars on a common substrate but interconnected through channel nodes with small resistances.

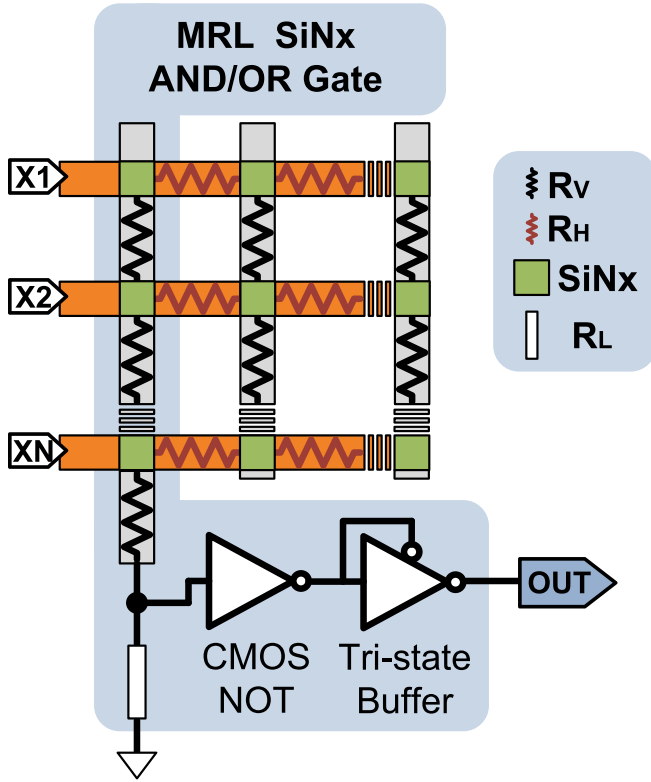

**Fig. 5.** A diagram showing how the MRL gate can be implemented into a crossbar circuit architecture.

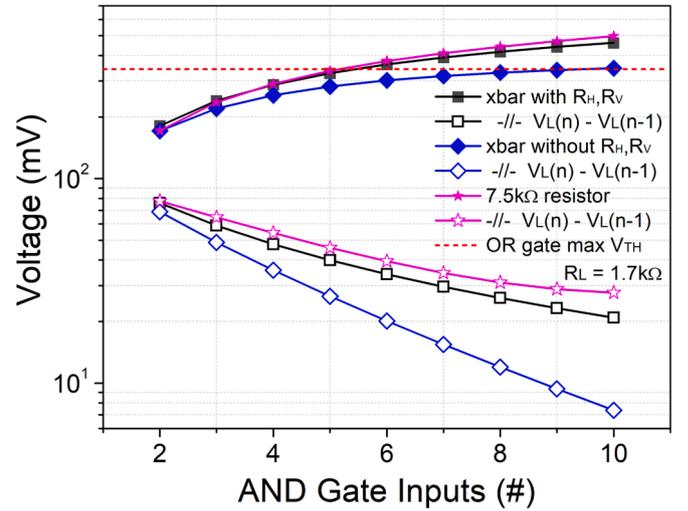

**Fig. 6.** SPICE simulations for MRL AND gates with different input size considering intermediate crossbar resistances.

Finally, power efficiency considerations were made for the proposed MRL structure. In Fig. 7(a), the power consumption for different input sizes for the MRL AND/OR gates is shown. An asymptotic increment in power appears, proportional to the increase in the current of each crossbar column that implements the corresponding MRL gate. As expected, the power is increased due to the low resistances that have been chosen for the demonstration of the proposed MRL structure in order to maintain the validity of the fitting model as presented in Eq. (1). For this to happen must be $|X-V_L| = [0.4, 1.1]$ V. Under this condition we have performed another simulation for 2-input MRL AND/OR gates for the same range of conductance states $S = [1,12]$ and resistance $R_L = 200k\Omega$. As shown in Fig. 7(b) the power decreased by 3 orders of magnitude. Even more power reduction could be achieved by reducing the input voltage, which, however, in order to be simulated, the model would need to be readjusted to low voltage measurements.

## 4. Conclusions

This research effort involved the fabrication and characterization of a crossbar array of silicon nitride resistive memory on a silicon-on-insulator wafer. A dedicated fitting model was used to show the capability of multi-level-cell operation and 12 clearly distinct levels were identified. A hybrid silicon nitride RRAM crossbar and CMOS reconfigurable logic integration scheme based on memristor rationed logic was suggested and several circuit aspects were simulated and discussed. In an extension of this work, a specific tuning protocol for selector-less SOI silicon nitride RRAM crossbar, a silicon nitride low voltage I-V fitting model and an interconnection methodology with a segmented crossbar array architecture is currently under investigation.

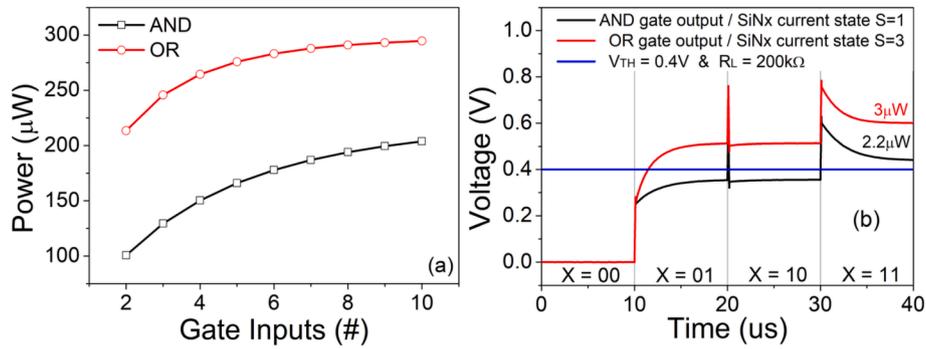

**Fig. 7.** SPICE simulations for power consumption estimation for different input sizes for the MRL AND/OR gates.

**Declaration of Competing Interest**

The authors declare the following financial interests/personal relationships which may be considered as potential competing interests: N. Vasileiadis reports financial support was provided by Hellenic Foundation for Research and Innovation. P. Dimitrakis reports financial support was provided by Hellenic Foundation for Research and Innovation. A. Mavropoulis reports financial support was provided by Hellenic Foundation for Research and Innovation.

**Data availability**

Data will be made available on request.

**Acknowledgements**

This work was supported by the research project "LIMA-chip" (Proj. No. 2748) funded by the Hellenic Foundation of Research and Innovation (HFRI).